\documentclass{article}
\usepackage{ijcai25}

\usepackage{times}
\usepackage{natbib}
\usepackage{soul}
\usepackage{url}
\usepackage[hidelinks]{hyperref}
\usepackage[utf8]{inputenc}
\usepackage[small]{caption}
\usepackage{graphicx}
\usepackage{amsmath}
\usepackage{amsthm}
\usepackage{booktabs}
\usepackage{algorithm}
\usepackage{algorithmic}
\usepackage[switch]{lineno}
\usepackage{colortbl} 
\usepackage{tikz}
\usetikzlibrary{bayesnet}
\usepackage{algorithm}
\usepackage{algorithmic}

\usepackage{fancyhdr}

\title{PEARL: Private Equity Accessibility Reimagined with Liquidity} 

\author{
E. Benhamou$^{1,2}$ \and
JJ. Ohana$^1$ \and
B. Guez$^1$ \and
E. Setrouk$^1$ \and
T. Jacquot$^1$ \and
\affiliations
$^1$Ai for Alpha\\
$^2$Dauphine PSL\\
\emails
\{eric.benhamou, jean-jacques.ohana, beatrice.guez, ethan.setrouk, thomas.jacquot\}@aiforalpha.com}

\begin{document}
\maketitle

\begin{abstract}
In this work, we introduce PEARL (Private Equity Accessibility Reimagined with Liquidity), an AI-powered framework designed to replicate and decode private equity funds using liquid, cost-effective assets. Relying on previous research methods such as Erik Stafford's single stock selection (Stafford) and Thomson Reuters - Refinitiv’s sector approach (TR), our approach incorporates an additional asymmetry to capture the reduced volatility and better performance of private equity funds resulting from sale timing, leverage, and stock improvements through management changes. As a result, our model exhibits a strong correlation with well-established liquid benchmarks such as Stafford and TR, as well as listed private equity firms (Listed PE), while enhancing performance to better align with renowned quarterly private equity benchmarks like Cambridge Associates, Preqin, and Bloomberg Private Equity Fund indices. Empirical findings validate that our two-step approach—decoding liquid daily private equity proxies with a degree of negative return asymmetry—outperforms the initial daily proxies and yields performance more consistent with quarterly private equity benchmarks.
\end{abstract}

\section{Introduction}\label{sec:Introduction}
Over the past quarter-century, investments in private equity (PE) have significantly increased in size, as illustrated by various strategic consulting companies reports such as  \cite{bcg2020global}, \cite{mckinsey2020annual} and lately  \cite{McKinsey2023}. As of June 30, 2023, private equity funds' assets under management (AUM) totaled over \$13 trillion, reflecting a nearly 2\% annual growth rate since 2018 as illustrated in  \cite{McKinsey2023}. Large institutional investors, including pension funds and sovereign wealth funds, allocate substantial portions of their portfolios to private investments. For instance, the California Public Employees' Retirement System (CalPERS) has increased its target exposure to private markets from 33\% to 40 \%, committing approximately \$15 billion to co-investments over the past 18 months. Similarly, Singapore's GIC, a sovereign wealth fund managing over \$700 billion in assets, actively invests in private equity, including acquiring stakes in Western companies' Chinese units, with undisclosed amounts rumoured though to be substantial.  Likewise, as of June 30, 2024, Harvard University's endowment (respectively  the Canada Pension Plan Investment Board (CPPIB) and the Yale University's endowment) was reported to have invested 39\% (respectively 32\% and 33 \%) of its \$53.2 billion  (respectively \$143 billion and \$40 billion) capital.

However despite the capacity of PE funds to offer the potential for much higher returns than public equities markets, one of the major constraint is their inherent illiquidity. Investments in PE involve investing in illiquid, privately held assets that requires long-term capital commitments. 
This illiquidity distinguishes PE from public equities and introduces substantial challenges for investors, particularly during periods with limited exit opportunities. During such times, the illiquidity of PE becomes especially pronounced and burdensome for large institutional investors who critically need to liquidate portions of their portfolio to fulfill obligations, such as distributions to pensioners. 

This has led to strong interest in developing liquid alternatives to PE funds that can reproduce similar performance streams while avoiding the constraints of illiquidity beyond evergreen solutions that are very constrained by the size of the secondary market. However, the feasibility of such alternatives remains uncertain in the economic literature. It is often argued that the superior returns associated with PE funds are intrinsically tied to the illiquidity of their investment portfolios. Investors are generally willing to commit their capital only when they anticipate higher returns as compensation for bearing illiquidity risks as argued in \cite{Kaplan_2005, Ang2014, ang2014estimating, Franzoni2012}.

There has been various attempts to offer liquid replicating benchmark. \cite{Stafford2021} uses small stocks selection while \cite{thomson_reuters_pe_index} uses a sector based approach that provides greater scalability by focusing on broad industry groups. While these approaches address some liquidity concerns, they may fail first to be scalable as they involve public though not very liquid or scalable assets and second to capture the nuanced value drivers in private equity, including the impact of management changes, operational improvements, and the timing of strategic sales, as highlighted by \cite{Kaplan2009leveraged}, resulting in some lower volatility and better performance in downturn times. The key question is whether we could use more liquid instruments like large equity indexes futures that can scale to a much higher extent. Additionally, incorporating asymmetry in returns has emerged as a critical innovation. Studies such as \cite{Ang2014} and \cite{Franzoni2012} highlight that private equity returns often exhibit asymmetric characteristics due to their exposure to risk-off environments and tail risk strategies. These factors contribute to the positive alpha observed in PE funds, making it imperative to account for these dynamics in any replication framework. By integrating these asymmetric return factors, hybrid strategies can better mimic the risk-return profile of private equity, offering a more comprehensive solution to liquidity constraints.

\subsection{Innovation}\label{sec:Innovation}
Our contributions to the field of replicating private equity funds are threefold:

\begin{itemize}
    \item \textbf{Enhanced Liquidity:} We achieve greater liquidity by using highly liquid futures contracts, offering a scalable and practical solution for replicating private equity returns in public markets.
    
    \item \textbf{Improved Replication Precision:} By employing advanced machine learning decoding techniques based on graphical models, which enhance traditional linear regression methods, our approach delivers more accurate replication of private equity performance.
    
    \item \textbf{Incorporation of Asymmetry in Decoding:} Our method accounts for the asymmetry in the replication process to address the embedded and implicit put option arising from the ability to time sales as well as other private equity funds performance enhancement like management improvement and capital optimization, ensuring robustness in performance replication.
\end{itemize}

\subsection{Structure of Paper}
The rest of the paper is organized as follows:  
Section~\ref{sec:Literature Review} reviews the existing literature on private equity and replication strategies. It begins by exploring the three pillars of value creation in private equity: corporate governance, operational optimization, and financial engineering. Additionally, this section discusses attempts to provide liquid replication strategies for private equity returns, evaluating their methodologies, performance, and limitations.

Section~\ref{sec:Methodology} presents the methodological framework adopted in this study. It begins with a description of the two-step approach used to analyze private equity replication strategies, providing a systematic breakdown of the process. We also provide a primer on graphical models to explain why these models generalize and improve advanced successive linear regressions or Kalman filters. Furthermore, the section discusses the role of asymmetric factors in capturing unique aspects of private equity performance and risk-return profiles.

Section~\ref{sec:Results and statistics} provides the empirical results and statistical analyses. This section evaluates the impact of benchmark selection on the performance of replication strategies, focusing on how different benchmarks slightly change the results. Additionally, the correlation accuracy of the proposed replication models is analyzed to assess their robustness in mimicking private equity performance under various conditions.

Finally, in the concluding section~\ref{sec:Conclusion}, we summarize key findings and discuss the limitations of the study, and suggest directions for future research.

\section{Literature review\label{sec:Literature Review}}
\subsection{Private Equity value creation}
Private equity (PE) funds differ significantly from venture capital in their investment strategy, focusing on acquiring majority control of mature firms rather than investing in emerging companies without majority control. The three pillars of value creation in leveraged buyouts—corporate governance, operational optimization, and financial engineering—have been central to PE value generation as explained in \cite{Kaplan2009leveraged}. The authors emphasize that these pillars enable general partners to introduce disciplined management incentives, optimize operations, and enhance financial structures within their portfolio companies.

Management incentivization is a cornerstone of private equity governance. For instance, \cite{Kaplan_2005} found that transitioning firms from public to private ownership significantly increases management equity stakes, aligning their incentives with long-term growth objectives. Furthermore, the illiquid nature of PE compensation, through stock and options, ensures a focus on sustained value creation over short-term manipulation. \cite{acharya2009private} corroborate this finding, showing that PE-backed CEOs and management teams receive substantial equity stakes, reinforcing alignment between managers and investors.

Financial engineering is another critical value-creation mechanism in PE. By optimizing capital structures and leveraging debt, PE funds capitalize on tax shields and enhance returns. \cite{Jensen1986agency} highlights how leverage mitigates free cash flow problems by enforcing financial discipline. However, excessive leverage can increase financial distress risks, as evidenced by \cite{axelson2013borrow}, who found a negative relationship between extreme leverage and fund returns.

Operational optimization has gained increasing importance since the late 1980s. Top-performing PE firms now focus on sector-specific expertise and employ specialized professionals to implement productivity improvements, strategic changes, and cost-cutting measures. Studies show that PE-backed companies outperform their public counterparts in terms of operational efficiency, with cash flow-to-sales ratios improving significantly post-buyout as illustrated in \cite{Kaplan_2005} and \cite{acharya2009private}.

The risk-return profile of private equity is another distinguishing feature of the asset class. Traditional portfolio theory posits that asset returns are commensurate with their inherent risks \cite{Markowitz1952portfolio}. However, PE assets deviate from this paradigm due to their illiquid nature, unique risk exposures, and systematic risk factors absent in public markets. \cite{doskeland2018evaluating} argue that illiquidity necessitates a premium return, particularly during periods of market distress when PE holdings are difficult to liquidate. Moreover, diversification across vintage years can mitigate idiosyncratic risk, as shown by \cite{Robinson2016}.

\subsection{Attempt to provide liquid replicating strategies}
In the economic literature, there have been efforts to develop liquid replicating benchmarks for private equity (PE) returns. Replication strategies have gained attention as a means of emulating the performance of private equity investments through public market instruments, addressing challenges such as the opaque nature, irregular cash flows, and long-term commitment associated with PE investments.

\citet{rasmussen2015leveraged} were among the first to explore the replication of private equity performance by focusing on small, value-oriented, and leveraged public stocks. They developed a ranking system prioritizing smaller, cheaper stocks with above-median leverage and incorporated metrics such as debt paydown and improving asset turnover. Their annual top-25 stock portfolio demonstrated alphas of 9.56\% and 13.06\% under the CAPM and the Fama-French 3-factor model, respectively, further enhanced by liquidity and momentum factors. This approach emphasized disciplined, long-term investment strategies while acknowledging the higher volatility inherent in such portfolios, which is often masked in private equity due to infrequent valuations.

Building on this foundation, \cite{Stafford2021} introduced a strategy that replicates private equity buyouts by constructing portfolios of small public companies with low valuation multiples, such as EV/EBITDA. By applying incremental leverage and matching the holding periods of private equity investments, this strategy successfully simulates the risk-return profile of PE funds. Stafford’s analysis revealed alphas ranging from -2.4\% to 11\%, with the highest alpha observed in portfolios comprising stocks with the lowest valuations. This study demonstrated the feasibility of achieving competitive returns without the illiquidity constraints of private equity. However, replication portfolios face limitations, including irregular and infrequent cash flows and potential data inaccuracies, complicating risk-adjusted performance evaluation.

An alternative approach to replication is provided by Thomson Reuters Private Equity Index (as presented in \cite{thomson_reuters_pe_index}), which uses a sector-based methodology to achieve greater scalability. This method focuses on broad industry groups to create a replicable benchmark for private equity returns. By emphasizing sector allocation rather than individual company selection, this index offers a more scalable and accessible approach for investors seeking PE-like exposure. However, it may lack the granularity of strategies that rely on specific valuation multiples or leverage criteria

\cite{gupta2021valuing} provide a complementary perspective by introducing a strip-by-strip valuation method that constructs a replicating portfolio using cash flows from listed equities and fixed-income instruments. Their model splits the risk-adjusted returns of private equity investments into individual cash flow components, capturing the timing, risk, and macroeconomic correlations of PE returns. While not explicitly a replication strategy, this approach provides valuable insights into the granular drivers of private equity performance and highlights the limitations of traditional replication methods.

Despite the promise of replication strategies, several challenges remain. Private equity funds are characterized by irregular and discretionary cash flows, making it difficult to match their performance with public market portfolios. Traditional risk estimation methods, such as time-series regressions, often struggle to account for the lagged and non-transparent valuations of private equity funds. Furthermore, replicating the operational and governance improvements brought by private equity general partners remains a significant hurdle, as these value-creation mechanisms are difficult to mirror in public market portfolios.

\section{Methodology\label{sec:Methodology}}
\subsection{Goal}
The objective of this study is to replicate the performance of private equity benchmarks such as the Cambridge Associates, Preqin, or Bloomberg Private Equity Buyout Index. These benchmarks exhibit significantly higher Sharpe ratios, typically around 1.5, far above the traditional Sharpe ratio of approximately 0.5 observed in major large equity indexes. Moreover, they exhibit reduced drawdowns, lower volatility and superior annualized performance, ranging from 11\% to close to 15\%. However, these traditional benchmarks present significant limitations: they are published quarterly, often with delays of up to three months, making direct daily replication practically unfeasible.

\subsection{Performance of Private Equity Benchmarks}

Table~\ref{tab:quaterly benchmarks} presents the historical performance of two widely recognized private equity benchmarks: the Cambridge Associates (CA) and Preqin indexes. Both exhibit strong annualized returns (13.9\% and 14.2\%, respectively) and moderate volatility (8.9\% and 7.5\%). Their Sharpe ratios of 1.56 and 1.89 indicate a favorable risk-adjusted return profile, while their drawdowns remain relatively contained. Notably, the 10\% worst drawdowns are significantly lower than those observed in traditional equity markets, reinforcing the stability of these benchmarks.

\begin{table}[htbp]
  \centering
  \caption{Performance of Traditional Private Equity Indexes
  \label{tab:quaterly benchmarks}}
  \resizebox{\columnwidth}{!}{%
    \begin{tabular}{llr}
    \toprule
           & \textbf{Cambridge Associates (CA)} & \textbf{Preqin} \\
    \midrule
    Start Date & 31/03/2011 & 31/03/2011 \\
    End Date & 29/12/2023 & 29/12/2023 \\
    Annual Return & 13.9\% & 14.2\% \\
    Annual Volatility & 8.9\% & 7.5\% \\
    Skew   & -0.27  & 0.06 \\
    Kurtosis & 1.64   & 1.46 \\
    Sharpe Ratio & 1.56   & 1.89 \\
    Sortino Ratio & 2.18   & 2.66 \\
    Max DD & 9.5\% & 7.3\% \\
    10\% Worst DD & 3.7\% & 1.7\% \\
    Return/maxDD & 1.5    & 1.9 \\
    Return/Worst 10\% DD & 3.8    & 8.5 \\
    Sampling & quarterly & quarterly \\
    \bottomrule
    \end{tabular}%
  }
\end{table}

Similarly, the Bloomberg Private Equity indexes, shown in Table~\ref{tab:bloomberg}, illustrate another perspective on private equity performance. The Bloomberg PEALL index has a slightly lower return (11.4\%) compared to the Cambridge Associates and Preqin indexes but benefits from the lowest volatility (5.9\%) and the highest Sharpe ratio (1.95). The Bloomberg PEBUY index, on the other hand, delivers a stronger annual return of 13.2\% with a slightly higher volatility of 7.6\%. These indexes confirm the robust risk-return characteristics of private equity investments, with both outperforming traditional equity benchmarks on a risk-adjusted basis.

\begin{table}[htbp]
  \centering
  \caption{Performance of Bloomberg Private Equity Indexes\label{tab:bloomberg}}
  \resizebox{\columnwidth}{!}{%
    \begin{tabular}{llr}
    \toprule
           & \textbf{Bloomberg PEALL} & \textbf{Bloomberg PEBUY} \\
    \midrule
    Start Date & 31/03/2011 & 31/03/2011 \\
    End Date & 29/12/2023 & 29/12/2023 \\
    Annual Return & 11.4\% & 13.2\% \\
    Annual Volatility & 5.9\% & 7.6\% \\
    Skew   & -0.29  & -0.54 \\
    Kurtosis & 0.82   & 1.75 \\
    Sharpe Ratio & 1.95   & 1.73 \\
    Sortino Ratio & 4.07   & 2.34 \\
    Max DD & 5.2\% & 8.9\% \\
    10\% Worst DD & 2.\% & 2.1\% \\
    Return/maxDD & 2.2    & 1.5 \\
    Return/Worst 10\% DD & 5.6    & 6.2 \\
    Sampling & quarterly & quarterly \\
    \bottomrule
    \end{tabular}%
  }
\end{table}%

The correlation matrix in Table~\ref{tab:quarterly benchmark correlations} reveals two distinct benchmark groups. The Cambridge Associates (CA) benchmark is relatively independent, with a 61\% correlation with Preqin and even lower correlations with the Bloomberg indexes (PEBUY: 75\%, PEALL: 70\%), suggesting a unique return pattern.

In contrast, Preqin, PEBUY, and PEALL are highly correlated, with Preqin at 91\% with PEBUY and 96\% with PEALL, while PEBUY and PEALL are nearly identical at 97\%. This suggests that Preqin closely tracks the Bloomberg indexes, likely due to similar methodologies.

Thus, the benchmarks form two distinct families: (1) the independent CA benchmark and (2) the highly correlated Preqin, PEBUY, and PEALL benchmarks.

\begin{table}[htbp]
  \centering
  \caption{Correlation between quarterly benchmarks \label{tab:quarterly benchmark correlations}}
  \resizebox{0.75 \columnwidth}{!}{%
    \begin{tabular}{|c|rrrr}
    \toprule
           & \multicolumn{1}{c}{\textbf{CA}} & \multicolumn{1}{c}{\textbf{Prequin}} & \multicolumn{1}{c}{\textbf{PEBUY}} & \multicolumn{1}{c}{\textbf{PEALL}} \\
    \midrule
    \textbf{CA} & 100\%  &        &        &  \\
    \textbf{Prequin} & 61\%   & 100\%  &        &  \\
    \textbf{PEBUY} & 75\%   & 91\%   & 100\%  &  \\
    \textbf{PEALL} & 70\%   & 96\%   & 97\%   & 100\% \\
    \cmidrule{1-1}    
  \end{tabular}
  }
\end{table}%

\subsection{Challenges and Alternatives with daily indexes}

Despite their strong performance, these quarterly benchmarks suffer from reporting delays and lack of daily liquidity, making them impossible to replicate in real-time portfolios. As an alternative, daily liquid proxies such as the Erik Stafford (Stafford) single-stock selection method implemented via the SummerHaven Private Equity Strategy Index (SHPEISM Index), the Thomson Reuters Refinitiv (TR) Private Equity Benchmark provide more frequent quotations.

Table~\ref{tab:daily_index} highlights the key differences between these daily liquid indexes and traditional private equity benchmarks. While the Thomson Reuters index achieves a respectable 12.5\% annualized return, its volatility (24.8\%) is significantly higher than that of quarterly benchmarks. Similarly, the Erik Stafford method, though often cited as a viable daily proxy, delivers lower annualized returns (10.9\%) and exhibits even greater volatility (25.9\%). The Listed Private Equity (Listed PE) index, while sharing the same annualized return (10.9\%) as the Stafford index, demonstrates lower volatility (20.2\%), resulting in a slightly improved Sharpe ratio (0.54). However, it exhibits a highly negative skew (-0.78) and extreme kurtosis (18.2), indicating a tendency toward large outlier events. This suggests that while the Listed index may appear more stable under normal market conditions, it is more prone to extreme tail-risk events. Its maximum drawdown (50.4\%) is the highest among the three indexes, further reinforcing this concern. Interestingly, the \textbf{10\% worst drawdowns} (24.8\%) are lower than those observed for the TR index (33.4\%), suggesting that its typical downside risk is more contained despite its vulnerability to extreme tail losses.

\begin{table}[htbp]
  \centering
  \caption{Performance of Daily Liquid Indexes \label{tab:daily_index}}
  \resizebox{\columnwidth}{!}{%
    \begin{tabular}{lrrr}
    \toprule
           & \textbf{Stafford} & \textbf{TR}     & \textbf{Listed PE} \\
    \midrule
    \midrule
    Start Date & 31/03/2011 & 31/03/2011 & 31/03/2011 \\
    End Date & 21/01/2025 & 21/01/2025 & 21/01/2025 \\
    Annual Return & 10.90\% & 12.50\% & 10.90\% \\
    Annual Volatility & 25.90\% & 24.80\% & 20.20\% \\
    Skew   & -0.33  & -0.64  & -0.78 \\
    Kurtosis & 3.02   & 1.55   & 18.2 \\
    Sharpe Ratio & 0.42   & 0.5    & 0.54 \\
    Sortino Ratio & 0.52   & 0.62   & 0.63 \\
    Max DD & 47.20\% & 41.70\% & 50.40\% \\
    10\% Worst DD & 21.10\% & 33.40\% & 24.80\% \\
    Return/maxDD & 0.2    & 0.3    & 0.2 \\
    Return/Worst 10\% DD & 0.5    & 0.4    & 0.4 \\
    Sampling & daily  & daily  & daily \\
    \bottomrule
    \end{tabular}%
  }%
\end{table}%

As a result, these three daily benchmarks (Stafford, TR and Listed PE) suffer from substantially lower Sharpe ratios (0.42, 0.5 and respectively 0.54) and much deeper drawdowns (max drawdowns of 47.2\%, 41.7\% and respectively 50.4\%). These findings underscore the challenges in replicating private equity performance with daily liquid instruments, as the trade-off between liquidity and performance (both in terms of Sharpe ratio and maximum drawdowns) remains substantial.

\subsection{Objective of This Study}

Given these limitations, this study aims to bridge the gap between quarterly private equity benchmarks and daily tradable alternatives. The goal is to construct a daily-replicable strategy that maintains the attractive risk-return profile of private equity indexes while avoiding their structural drawbacks, such as reporting delays and illiquidity. By leveraging insights from traditional and liquid proxies, this approach seeks to create an optimized model capable of closely mimicking private equity performance with a higher frequency of valuation.

\subsection{Two-step Approach}
To achieve this ambitious goal, we employ a two-step approach. First, we decode the daily benchmark with high accuracy using advanced graphical models, as outlined in Section~\ref{sec:Graphical model primer}. This approach improves upon traditional replication methods, such as repetitive penalized linear regression, which suffers from issues like inconsistent weights and difficulties in determining when to update the model. \cite{roncalli2007alternative} explored Kalman filtering to provide dynamic updates and correction mechanisms. Building on this, \cite{benhamou2024grip}, \cite{Ohana2022deep} demonstrated that incorporating graphical models could enhance Kalman filters by introducing richer interaction modeling between assets.

The second step introduces asymmetries to address the discrepancies between daily benchmarks and quarterly private equity indexes. By incorporating tailored factors, we aim to replicate the more stable performance profiles of quarterly benchmarks while capturing critical dynamics such as drawdown mitigation and volatility control. 
The asymmetrical transformation of the daily index accounts for the tendency of private equity funds to mitigate negative returns. This is achieved by scaling down negative returns to reduce drawdown, mimicking the arbitrary valuation practices often employed by private equity funds. Specifically, for any negative returns, only \(9\%\) of the return is considered. Additionally, we introduce new asymmetrical factors, denoted as \(AF\), to represent this adjustment.

The initial daily private equity benchmark return \(R_t\) is transformed as follows:

\[
R_t' =
\begin{cases}
R_t, & \text{if } R_t \geq 0, \\
AF \cdot R_t, & \text{if } R_t < 0,
\end{cases}
\]

where \(AF = 0.9\). 

Thus, the transformation ensures that positive returns remain unchanged, while negative returns are scaled down to \(9\%\), reflecting the drawdown adjustment characteristic of private equity valuations.

The full methodology is outlined in Algorithm~\ref{alg:decoding_proxy}, following a structured process with the different steps described below:

\begin{itemize}
    \item We first decode a liquid daily proxy of private equity funds based either on the Erik Stafford, Thomson Reuters or he listed private equity indices (step 1).
    \item We then introduce asymmetry in the returns by applying an adjustment factor to account for reduced drawn downs to better align with quarterly private equity funds indexes (step 2)
    \item We finally evaluate the performance of decoding in terms of returns compared to quarterly benchmark (steps 3 and 4).
    \item We finally add some model constraints to ensure the model will stay within historical bounds and to prevent extreme weight deviations (steps 5 and 6).
\end{itemize}

\begin{algorithm}
\caption{Decoding Liquid Daily Proxy for Private Equity Funds with Additional Steps for Sanity Check}
\label{alg:decoding_proxy}
\begin{algorithmic}[1]
\STATE \textbf{Step 1:} Decode liquid daily proxy for private equity funds using equity index futures either using Erik Stafford, Thomson Reuters or the S\&P listed private equity index

\STATE \textbf{Step 2:} Introduce asymmetry in decoding the daily benchmark by reducing the size of negative returns by ten percents.

\STATE \textbf{Step 3:} Perform Maximum Likelihood Optimization on model parameters during training to generate the initial decoding model

\STATE \textbf{Step 4:} Run backtesting only on the test period with continuous prediction correction steps

\STATE \textbf{Step 5:} Check historical weights as a sanity check and validate that weights are not blowing up.
\STATE \textbf{Step 6:} Add constraints on the minimum and maximum historical weights to prevent model explosion
\end{algorithmic}
\end{algorithm}

\subsection{Graphical Model Primer\label{sec:Graphical model primer}}
Graphical models provide a robust framework for dynamic Bayesian inference, enabling us to decode daily benchmarks effectively. These models represent the relationships between variables as a network of nodes (variables) and edges (dependencies), facilitating the estimation of the most probable allocations at each time step. 

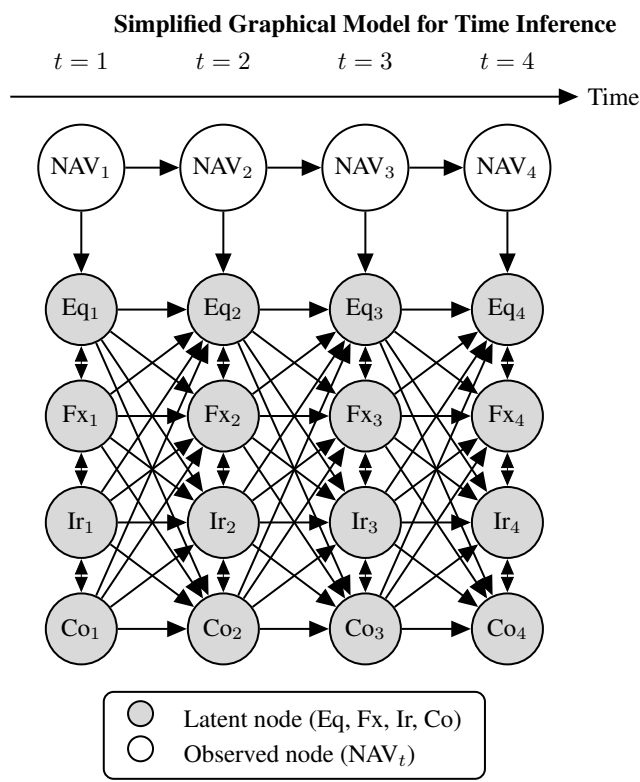
\begin{figure}[ht]
    \centering
  \resizebox{\columnwidth}{!}{%
    \begin{tikzpicture}
        \tikzstyle{latent} = [circle, fill=gray!30, minimum size=1cm, draw=black, thick]
        \tikzstyle{observed} = [circle, fill=white, minimum size=1cm, draw=black, thick]
        \tikzstyle{legendbox} = [rectangle, draw=black, rounded corners, thick, fill=white]

        \draw[thick, ->] (-1, 0) -- (7, 0) node[right] {Time};

        \foreach \x\y in {0/1, 2/2, 4/3, 6/4}{
            \node at (\x, 0.5) {$t=\y$};
        }

        \foreach \x\y in {0/1, 2/2, 4/3, 6/4} {
            \node[observed] (NAV\x) at (\x, -1) {NAV$_\y$};
        }

        \foreach \x\y in {0/1, 2/2, 4/3, 6/4} {
            \node[latent] (Eq\x) at (\x, -3) {Eq$_\y$};

            \node[latent] (Fx\x) at (\x, -4.5) {Fx$_\y$};

            \node[latent] (Ir\x) at (\x, -6) {Ir$_\y$};

            \node[latent] (Co\x) at (\x, -7.5) {Co$_\y$};
        }

        \foreach \x/\y in {0/2, 2/4, 4/6} {
            \draw[->, thick] (NAV\x) -- (NAV\y);
        }

        \foreach \x/\y in {0/2, 2/4, 4/6} {
            \draw[->, thick] (Eq\x) -- (Eq\y);
            \draw[->, thick] (Eq\x) -- (Fx\y);
            \draw[->, thick] (Eq\x) -- (Ir\y);
            \draw[->, thick] (Eq\x) -- (Co\y);

            \draw[->, thick] (Fx\x) -- (Eq\y);
            \draw[->, thick] (Fx\x) -- (Fx\y);
            \draw[->, thick] (Fx\x) -- (Ir\y);
            \draw[->, thick] (Fx\x) -- (Co\y);

            \draw[->, thick] (Ir\x) -- (Eq\y);
            \draw[->, thick] (Ir\x) -- (Fx\y);
            \draw[->, thick] (Ir\x) -- (Ir\y);
            \draw[->, thick] (Ir\x) -- (Co\y);

            \draw[->, thick] (Co\x) -- (Eq\y);
            \draw[->, thick] (Co\x) -- (Fx\y);
            \draw[->, thick] (Co\x) -- (Ir\y);
            \draw[->, thick] (Co\x) -- (Co\y);
        }

        \foreach \x in {0, 2, 4, 6} {
            \draw[<-, thick] (Eq\x) -- (NAV\x);
            \draw[<->, thin] (Fx\x) -- (Eq\x);
            \draw[<->, thin] (Ir\x) -- (Fx\x);
            \draw[<->, thin] (Co\x) -- (Ir\x);
        }

        \node[legendbox] (legend) at (3, -9) {
            \begin{tabular}{ll}
            \tikz\draw[latent] (0,0) circle (5pt); & Latent node (Eq, Fx, Ir, Co) \\
            \tikz\draw[observed] (0,0) circle (5pt); & Observed node (NAV$_t$) \\
            \end{tabular}
        };

        \node at (4, 1) {\textbf{Simplified Graphical Model for Time Inference}};
    \end{tikzpicture}
    }
    \caption{Simplified graphical model showing the relationship between observed NAv and inferred allocation as time goes by. For illustration purpose, we use different assets, with one being an Equity shortened in Eq, a second one an exchange rate shorted in Fx, a third one, an interest rates instrument shortened in Ir and finally a commodity asset shortened in Co.}
    \label{fig:graphical_model}
\end{figure}

The methodology begins with constructing a probabilistic graphical model, as illustrated in Figure~\ref{fig:graphical_model}. The overall idea is to capture the relationships between the weights in the various assets and the observed NAVs. The  dynamic inference is done by incorporating time-dependent priors and observations, making it well-suited for daily updates. If we denote by $NAV_t$, the observed net asset value of a fund, and assume various assets (Eq, Fx, Ir and Co), we are interested in finding the weights in these assets given the observed NAVs. If we denote by $w^{eq}_t, w^{Fx}_t, w^{Ir}_t$ and  $w^{Co}_t$ the respective weights, and by $r^{eq}_t, r^{Fx}_t, r^{Ir}_t$ and  $r^{Co}_t$, as well as the predicted $\widehat{NAV}_t$, the underlying relationship between latent nodes and observed node is as follows:

\begin{equation}\label{eq:dynamic}
    \widehat{NAV}_t = \widehat{NAV}_{t-1} \left( 1+ w^{Eq}_{t-1} r^{Eq}_t + \ldots + w^{Co}_{t-1} r^{Co}_t \right)
\end{equation}

Equation~\ref{eq:dynamic} reflects that the estimated NAV at time \( t \) is obtained by adding the weighted returns of each asset to the previous estimated NAV, with the weights determined at the prior time step, \( t-1 \).

Key steps in the graphical model approach include:
\begin{enumerate}
    \item \textbf{State Space Representation}: The weights are modeled as latent states influenced by observable asset prices and previous weights. This model is more precise than the Kalman filter, as the Kalman filter only accounts for interactions within each individual asset, without considering interactions between different assets.
    \item \textbf{Dynamic Inference}: Bayesian inference techniques, such as message passing algorithms, is used to estimate the latent states at each time step with a prediction correction step like in Kalman filter, hence making it more robust than piecewise regressions that do not account for previous weights in their estimations.
    \item \textbf{Interaction Modeling}: The graphical model accounts for interactions between assets, allowing for richer modeling compared to traditional independent assumptions.
    \item \textbf{Updating Mechanisms}: The model dynamically updates the allocation probabilities as new data becomes available, ensuring adaptability to changing market conditions. This ensures that the model continuously improves itself.
\end{enumerate}

The graphical model framework enhances replication accuracy compared to other approach and provides insights into the underlying dynamics of the daily benchmark. Even if graphical models remain largely unknown to the financial community, they have been widely used in Machine Learning \cite{murphy2012machine} and have been the underlying building block for voice and world detection of the early versions of Apple Siri application.

\subsection{Asymmetric Factors}
To align the daily benchmark more closely with quarterly private equity indexes, we introduce asymmetric factors to address the discrepancies in drawdowns and risk profiles. Private equity benchmarks typically exhibit smoother performance and reduced volatility due to their infrequent valuations and long-term investment horizons. 

We incorporate the following asymmetric factors into our model:
\begin{enumerate}
    \item \textbf{Tail Risk Strategies}: Using volatility indices such as the VIX, we design strategies that hedge against extreme market movements, thereby mitigating pronounced drawdowns in the daily benchmark. A novel tail risk strategy is proposed to systematically capitalize on upward movements in VIX Futures by establishing long positions in either the front-end month (short-term, ST VIX) or the fourth-month (medium-term, MT VIX) futures, contingent upon prevailing market conditions. The core methodology employs a machine learning framework that identifies relatively infrequent yet pronounced positive trends in the VIX Futures curve based on three indicators: 
        \begin{enumerate}
            \item The 20-day Volatility-Adjusted Return on the VIX Future, which captures short-term momentum.
            \item The VIX Curve Future Ratio—defined as the ratio of the next VIX Future to the current VIX Future—to assess potential carry benefits for long VIX positions.
            \item The overall level of the VIX, which reflects the mean reversion dynamics that often arise during periods of heightened or protracted market volatility.
        \end{enumerate}
    Once these indicators generate a probabilistic “activation signal,” the model allocates positions to ST VIX and MT VIX accordingly. By toggling between the front-month and fourth-month segments of the VIX curve, this approach achieves a more nuanced exposure to volatility, allowing it to capture both acute market shocks and longer-duration volatility regimes.
    
    Empirical findings indicate that integrating this Tail Risk Hedge Vol Strategy as an overlay on equity-oriented portfolios can bolster risk-adjusted performance metrics. Over the period from January 2007 to December 2024, the strategy not only enhances returns but also reduces overall portfolio volatility, resulting in a 71\% increase in the Sharpe Ratio and a 2.5-fold improvement in the Return to Maximum Drawdown (Max DD). During low-volatility regimes—when equity markets tend to be complacent—this overlay provides a protective buffer against potential market downturns. Notably, the addition of the hedge lifts the annual return from 9\% to 12\%, while reducing annual volatility from 20\% to 16\% and curtailing the maximum drawdown from 56\% to 29\%. These results underscore the potential of a machine learning--driven tail risk hedge to enhance portfolio resilience and improve long-term investment outcomes.

    \item \textbf{Momentum Strategies}: Building on a traditional CTA cross-asset class framework, this strategy leverages a dedicated \emph{risk-off filter}, implemented through a non-linear Hysteresis Filtering Algorithm, to systematically identify and prioritize trends negatively correlated with the global equity market. Demonstrating robust diversification capabilities, the strategy achieves a -36\% correlation with the S\&P 500 and delivers positive returns in 88\% of months when the S\&P 500 declines by more than -5\% over the period 2010–2024. Furthermore, it generates an average monthly return of 3.6\% during such downturns, outperforming the Benchmark Index of CTAs by a factor of two. By integrating momentum-based factors with the risk-off filter, the strategy provides superior hedging performance during medium- to long-term equity market declines while maintaining consistent risk-adjusted returns. These attributes establish the strategy as a useful tool for mitigating portfolio drawdowns and enhancing resilience in volatile market environments.

\end{enumerate}

These asymmetric factors enhance the daily benchmark's performance by reducing drawdowns and aligning its risk-return profile more closely with quarterly benchmarks, making it a viable alternative for replicating private equity returns.

\section{Results and Statistics\label{sec:Results and statistics}}

\subsection{Data description}
We use for daily private equity funds three proxy indexes: the Erik Stafford whose Bloomberg ticker is SHPEI Index, the Thomson Reuters Refinitiv Private Equity Benchmark whose Bloomberg ticker is TRPEI Index and the S\&P listed private equity funds whose Bloomberg ticker is SPLPEQNT Index. We collect data from 2005 up to January 21st 2025. We train our graphical model from 2005 up to end of 2010. The test set starts from 31 March 2011 as the Preqin index only starts at this dates. The performance benchmark in terms of private equity funds are the seminal quarterly indexes used by the private equity industry, namely the Cambridge Associates, the Preqin  and the Bloomberg Private Equity Buyout (PEBUY) and the rivate Equity All (PEALL) Index.

\subsection{Replication Accuracy}
In order to determine the replication quality and accuracy in the proposed PEARL framework, we not only provide the key statistics in Table~\ref{tab:decoding stats} but also their respective correlation with their initial benchmarks in Table~\ref{tab:correlation with benchmarks}.

\begin{table}[htbp]
  \centering
  \caption{Performance of decoding of the three strategies: Listed Private Equities Companies (Listed PE), Thomson Reuters (TR) and Erik Stafford (Stafford)  \label{tab:decoding stats} }
  \resizebox{\columnwidth}{!}{%
    \begin{tabular}{lrrr}
    \toprule
           & \multicolumn{3}{c}{ \textbf{Decoding}}   \\
           &  \textbf{Listed PE} &  \textbf{TR} &  \textbf{Stafford}  \\
    \midrule
    \midrule
    Start Date & 02/01/2012 & 02/01/2012 & 02/01/2012 \\
    End Date & 21/01/2025 & 21/01/2025 & 21/01/2025 \\
    Annual Return & 17.1\% & 17.7\% & 17.4\% \\
    Annual Volatility & 13.6\% & 14.0\% & 14.2\% \\
    Auto Correlation & 3.5\% & 0.6\% & 1.0\% \\
    Skew   & -0.26  & -0.23  & -0.25 \\
    Kurtosis & 4.48   & 3.75   & 4.1 \\
    Sharpe Ratio & 1.26   & 1.27   & 1.23 \\
    Sortino Ratio & 1.69   & 1.72   & 1.67 \\
    Max DD & 17.6\% & 19.2\% & 19.2\% \\
    10\% Worst DD & 9.5\% & 8.6\% & 9.5\% \\
    Return/maxDD & 1      & 0.9    & 0.9 \\
    Return/Worst 10\% DD & 1.8    & 2.1    & 1.8 \\
    \end{tabular}%
    }
\end{table}%

\begin{table}[htbp]
  \centering
  \caption{Correlation of strategies with their benchmarks\label{tab:correlation with benchmarks}}
  \resizebox{\columnwidth}{!}{%
    \begin{tabular}{lrrrrrr}
    \toprule
    \multicolumn{1}{p{5em}}{\textbf{Strategy}} & \multicolumn{1}{p{4.145em}}{\textbf{Lifetime}} & \multicolumn{1}{p{1.93em}}{\textbf{1Y}} & \multicolumn{1}{p{1.93em}}{\textbf{3Y}} & \multicolumn{1}{p{1.93em}}{\textbf{5Y}} & \multicolumn{1}{p{1.93em}}{\textbf{7Y}} & \multicolumn{1}{p{2.285em}}{\textbf{10Y}} \\
    \midrule
    \midrule
    Stafford & 64\%   & 94\%   & 68\%   & 81\%   & 73\%   & 71\% \\
    Listed PE & 63\%   & 89\%   & 65\%   & 79\%   & 75\%   & 72\% \\
    TR     & 69\%   & 97\%   & 71\%   & 81\%   & 78\%   & 76\% \\
    \end{tabular}%
  }
\end{table}

Table~\ref{tab:decoding stats} presents the decoding performance of the three strategies: Listed Private Equities Companies (Listed PE), Thomson Reuters (TR), and Erik Stafford (Stafford). Across all strategies, the annualized returns are comparable, with TR achieving the highest return (17.7\%), followed closely by Stafford (17.4\%) and Listed PE (17.1\%). In terms of volatility, levels remain similar, with Listed PE exhibiting the lowest overall volatility (13.6\%) and Stafford the highest (14.2\%). 

The autocorrelation measure is highest for Listed PE (3.5\%), suggesting greater persistence in returns compared to TR (0.6\%) and Stafford (1.0\%). Skewness values indicate a slight negative bias across all strategies, with TR exhibiting the lowest absolute skew (-0.23), while kurtosis is highest for Listed PE (4.48), suggesting a higher occurrence of extreme events.

In terms of key performance ratios, as measured by the Sharpe and Sortino ratios, the three strategies are quite comparable with  TR displaying a slight edge in both metrics (Sharpe: 1.27, Sortino: 1.72). Drawdown analysis highlights that Listed PE has the lowest maximum drawdown (17.6\%) compared to TR and Stafford (both at 19.2\%). Finally, return-to-drawdown ratios indicate that Listed PE achieves the highest Return/MaxDD (1.0) and Return/Worst 10\% DD (1.8), although TR surpasses it in the latter metric (2.1). These findings suggest that while TR offers the highest returns and strong downside protection in extreme cases, Listed PE provides the most stable return profile with lower overall volatility and drawdown risk.

The results presented in Table~\ref{tab:decoding stats} (decoding performance of the PEARL strategy) demonstrate a significant improvement over the daily liquid benchmarks in Table~\ref{tab:daily_index}, while more closely aligning with the traditional private equity benchmarks in Tables~\ref{tab:quaterly benchmarks} and~\ref{tab:bloomberg}. 

First, the annualized returns of the PEARL strategy (17.1\%–17.7\%) are notably higher than those of the daily liquid indexes (10.9\%–12.5\%), while being comparable to, or exceeding, the traditional quarterly benchmarks (11.4\%–14.2\%). This suggests that the PEARL approach captures a return profile more consistent with long-term private equity performance.

Second, the Sharpe ratios for PEARL (1.23–1.27) significantly outperform those of the daily liquid indexes (0.42–0.54), highlighting superior risk-adjusted performance. While the Sharpe ratios remain slightly below those of traditional private equity benchmarks (1.56–1.95), the PEARL strategy represents a substantial enhancement in stability and efficiency relative to daily liquid proxies.

Additionally, PEARL exhibits lower maximum drawdowns (17.6\%–19.2\%) than the daily liquid benchmarks (41.7\%–50.4\%), reducing downside risk exposure. Although traditional private equity indices still demonstrate superior drawdown resilience, with max drawdowns below 10\%, PEARL significantly mitigates the volatility and deep drawdowns characteristic of daily replication strategies.

These findings confirm that the PEARL strategy achieves a better balance between liquidity and performance, bridging the gap between daily liquid instruments and traditional private equity benchmarks. By enhancing returns, reducing drawdowns, and improving risk-adjusted returns, PEARL represents a meaningful advancement over existing daily replication approaches.

\begin{table}[htbp]
  \centering
  \caption{Correlation between strategies}
  \label{tab:correlation_between_strategies}
  \resizebox{0.7 \columnwidth}{!}{%
    \begin{tabular}{|c|rrr}
    \toprule
           & \multicolumn{1}{c}{\textbf{Stafford}} & \multicolumn{1}{c}{\textbf{Listed PE}} & \multicolumn{1}{c}{\textbf{TR}} \\
    \midrule
    \textbf{Stafford} & 100\%  & 83\%   & 74\% \\
    \textbf{Listed PE} & 83\%   & 100\%  & 83\% \\
    \textbf{TR} & 74\%   & 83\%   & 100\% \\
    \end{tabular}%
    }
\end{table}%

\begin{table}[htbp]
  \centering
  \caption{Yearly returns  \label{tab:yearly returns}}
  \setlength{\tabcolsep}{3pt} 
  \resizebox{1 \columnwidth}{!}{%
    \begin{tabular}{crrrrrr}
    \toprule
           & \multicolumn{3}{c}{Benchmark} & \multicolumn{3}{c}{Decoding} \\
    \multicolumn{1}{l}{\textbf{Years}} & \multicolumn{1}{c}{\textbf{Listed PE}} & \multicolumn{1}{c}{\textbf{TR}} & \multicolumn{1}{c}{\textbf{Stafford}} & \multicolumn{1}{c}{\textbf{Listed PE}} & \multicolumn{1}{c}{\textbf{TR}} & \multicolumn{1}{c}{\textbf{Stafford}} \\
    \midrule
    \midrule
    \textbf{2025} & 4.5\%  & 4.6\%  & 3.2\%  & 2.2\%  & 1.8\%  & 1.6\% \\
    \textbf{2024} & 24.0\% & 31.3\% & 2.7\%  & 16.0\% & 16.1\% & 16.2\% \\
    \textbf{2023} & 39.0\% & 4.4\%  & 23.1\% & 22.5\% & 21.3\% & 20.6\% \\
    \textbf{2022} & \textcolor[rgb]{ 1,  0,  0}{-29.0\%} & \textcolor[rgb]{ 1,  0,  0}{-31.1\%} & \textcolor[rgb]{ 1,  0,  0}{-11.1\%} & \textcolor[rgb]{ 1,  0,  0}{-4.2\%} & \textcolor[rgb]{ 1,  0,  0}{-4.0\%} & \textcolor[rgb]{ 1,  0,  0}{-1.1\%} \\
    \textbf{2021} & 41.8\% & 29.8\% & 43.4\% & 13.2\% & 12.8\% & 10.3\% \\
    \textbf{2020} & 4.5\%  & 25.6\% & 17.1\% & 52.3\% & 53.2\% & 52.2\% \\
    \textbf{2019} & 44.6\% & 37.4\% & 12.6\% & 31.6\% & 32.0\% & 30.3\% \\
    \textbf{2018} & \textcolor[rgb]{ 1,  0,  0}{-14.0\%} & \textcolor[rgb]{ 1,  0,  0}{-11.9\%} & \textcolor[rgb]{ 1,  0,  0}{-14.7\%} & \textcolor[rgb]{ 1,  0,  0}{-8.9\%} & \textcolor[rgb]{ 1,  0,  0}{-7.9\%} & \textcolor[rgb]{ 1,  0,  0}{-8.4\%} \\
    \textbf{2017} & 24.3\% & 31.5\% & 7.5\%  & 13.8\% & 14.3\% & 12.2\% \\
    \textbf{2016} & 13.6\% & 8.6\%  & 28.5\% & 9.3\%  & 10.2\% & 11.2\% \\
    \textbf{2015} & \textcolor[rgb]{ 1,  0,  0}{-3.2\%} & 6.7\%  & \textcolor[rgb]{ 1,  0,  0}{-9.4\%} & 7.7\%  & 9.8\%  & 8.0\% \\
    \textbf{2014} & \textcolor[rgb]{ 1,  0,  0}{-1.6\%} & 20.6\% & 7.8\%  & 32.5\% & 39.6\% & 40.2\% \\
    \textbf{2013} & 35.3\% & 42.6\% & 47.0\% & 32.2\% & 30.7\% & 31.7\% \\
    \textbf{2012} & 29.0\% & 20.4\% & 17.9\% & 16.0\% & 15.4\% & 16.0\% \\
    \end{tabular}%
    }
\end{table}%

Table~\ref{tab:correlation_between_strategies} is also consistent with the findings that the three strategies are quite similar. It shows strong correlations among the three strategies, with Listed PE exhibiting the highest alignment (83\%) with both Stafford and TR. TR has the lowest correlation with Stafford (74\%), suggesting some diversification benefits. Overall, the high correlations indicate that the choice of the liquid daily benchmark is not so meaningful.

Additionally, Table~\ref{tab:correlation with benchmarks} demonstrates that the PEARL method achieves a high correlation with its decoded benchmarks, particularly in shorter time horizons. The one-year correlations exceed 89\% for all strategies, indicating strong short-term alignment with the benchmark. Over the lifetime period, correlations remain robust (63\%–69\%), confirming the PEARL method's ability to effectively replicate private equity characteristics while maintaining liquidity. These results highlight the strategy’s effectiveness in capturing the dynamics of private equity performance with high fidelity.

\begin{figure}[htbp]
  \centering
  \resizebox{\columnwidth}{!}{%
    \includegraphics{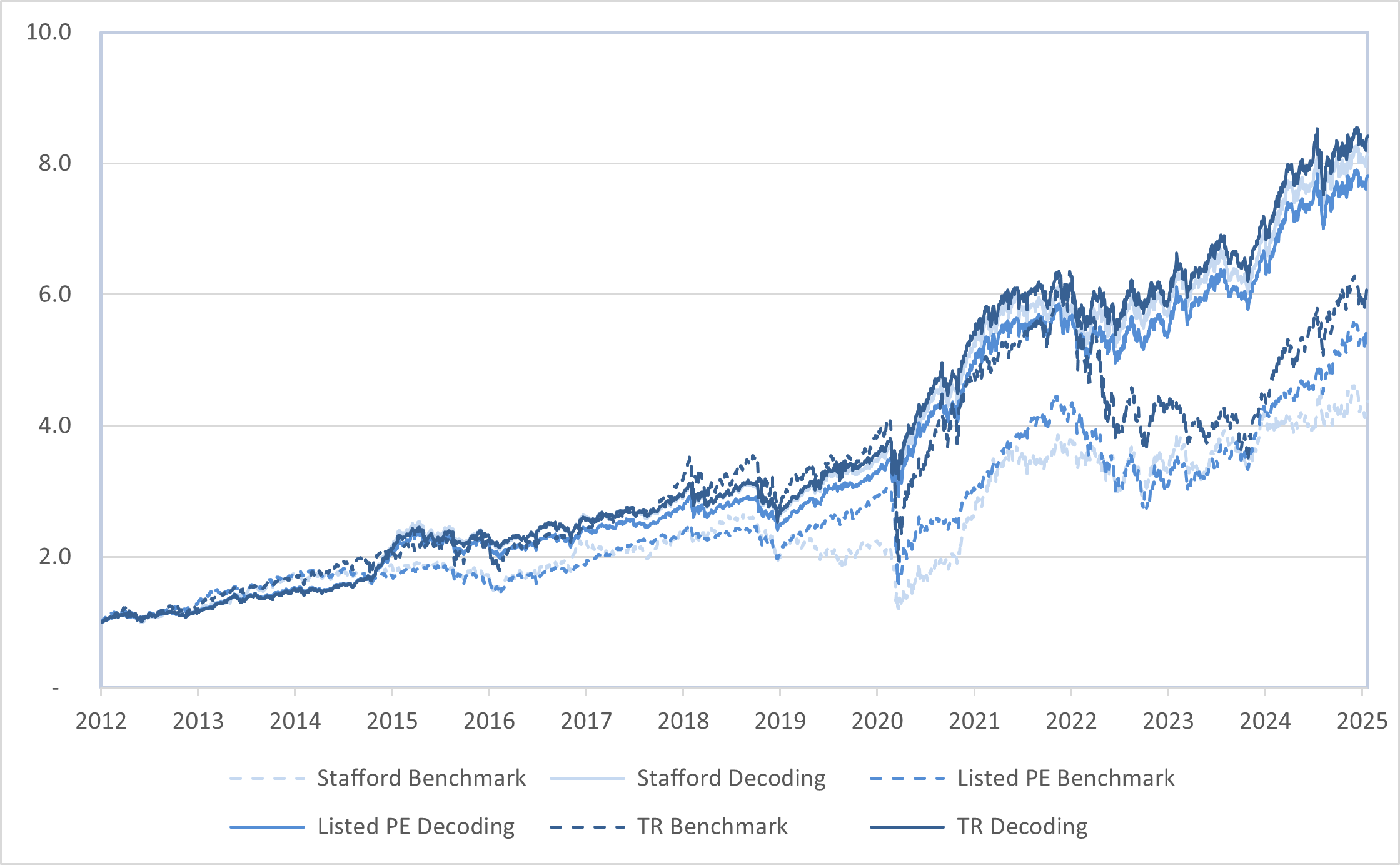}
  }
  \caption{Comparison of Private Equity Returns decoding strategies with their benchmarks}
  \label{fig:pe_comparison}
\end{figure}

Last but not least, Table~\ref{tab:yearly returns} and Figure~\ref{fig:pe_comparison} illustrate the yearly performance of private equity decoding strategies relative to their benchmarks. The PEARL method consistently tracks benchmark trends while mitigating extreme losses, particularly during downturns (e.g., 2022 and 2018). These results highlight the strategy's effectiveness in capturing private equity returns with improved downside protection.

Figure~\ref{fig:pe_comparison} visually compares the PEARL decoding strategies with their respective private equity benchmarks. The figure highlights the strong alignment between the decoded and benchmark returns while demonstrating the PEARL method’s ability to reduce volatility and mitigate extreme drawdowns and hence reproduce performance more in line with the quarterly benchmarks like the ones of Cambridge Associate, Preqin or Bloomberg indexes. Notably, the strategy effectively smooths negative shocks, as observed in 2022 and 2018, while maintaining competitive returns in high-growth periods, in a similar way as private equity funds do.

\section{Conclusion\label{sec:Conclusion}}
In this work, we introduced PEARL (Private Equity Accessibility Reimagined with Liquidity), an innovative AI-driven framework designed to replicate private equity (PE) funds using liquid and cost-effective assets. The methodology leverages advanced graphical models to decode daily benchmarks with high precision, addressing the limitations of existing replication approaches. By incorporating asymmetric factors such as tail risk and momentum strategies, PEARL aligns daily benchmark performance more closely with renowned private equity indexes like the Cambridge Associates and Preqin benchmarks. Overall, PEARL provides a significant step forward in making private equity funds liquid while maintaining its core performance characteristics in line with top private equity funds.
Future research could explore enhancements to the graphical model framework, particularly through deeper integration of macroeconomic indicators and alternative asset classes, to further improve replication accuracy.

\bibliographystyle{ijcai25}
\bibliography{main}

\end{document}